\documentclass[structabstract]{aa}

\usepackage{graphicx}
\usepackage{txfonts}

\begin{document}

   \title{Constructing a Galactic coordinate system based on near-infrared and radio catalogs}

   \author{J.-C. Liu \inst{1,2} \and
          Z. Zhu\inst{1,2}
                                        \and
             B. Hu\inst{3,4}
          }

   \institute{Department of astronomy, Nanjing University, Nanjing 210093, China\\
              \email{[jcliu;zhuzi]@nju.edu.cn}
         \and
             key Laboratory of Modern Astronomy and Astrophysics (Nanjing University), Ministry of Education, Nanjing 210093, China
            \and
                 Purple Mountain Observatory, Chinese Academy of Sciences, Nanjing 210008, China
          \and
                  Graduate School of Chinese Academy of Sciences, Beijing 100049, China\\
                    \email{hubo@pmo.ac.cn}
             }

   \date{Received ; accepted}
    \titlerunning{The optimal Galactic coordinate system}
    \authorrunning{Liu et al.}


  \abstract
   {The definition of the Galactic coordinate system was announced by the IAU Sub-Commission 33b on behalf of the IAU in 1958. An unrigorous transformation was adopted by the Hipparcos group to transform the Galactic coordinate system from the FK4-based B1950.0 system to the FK5-based J2000.0 system or to the International Celestial Reference System (ICRS). For more than 50 years the definition of the Galactic coordinate system has remained unchanged from this IAU1958 version. On the basis of deep and all-sky catalogs, the position of the Galactic plane can be revised and updated definitions of the Galactic coordinate systems can be proposed.}
   {We re-determine the position of the Galactic plane based on modern large catalogs, such as the Two Micron All-Sky Survey (2MASS) and the SPECFIND v2.0. This paper also aims to propose a possible definition of the optimal Galactic coordinate system by adopting the ICRS position of the Sgr A* at the Galactic center.}
   {The near-infrared 2MASS point source catalog and the SPECFIND v2.0 catalog of radio continuum spectra are used to determine the mean position of the Galactic plane on the celestial sphere. By fitting the data to an ideal Galactic equator, the parameters defining the Galactic coordinate system are obtained.}
   {We find that the obliquity of the Galactic equator on the ICRS principal plane is about $0.4^\circ$ (2MASS) and $0.6^\circ$ (SPECFIND v2.0) larger than the J2000.0 value, which is widely used in coordinate transformations between the equatorial $(\alpha, \delta)$ and the Galactic $(\ell, b)$. Depending on the adopted parameters, data, and methods, the largest difference between the resulting Galactic coordinate systems is several arcminutes. We derive revised transformation matrices and parameters describing the orientation of the Galactic coordinate systems in the ICRS at the 1 milli-arcsecond level to match the precision of modern observations.}
   {For practical applications, we propose that a revised definition of the Galactic coordinate system should be required in the near future.}


   \maketitle
%

\section{Introduction}

The Galactic coordinate system (hereafter GalCS) is important for
studies of the Galactic structure, kinematics, and dynamics. In the
GalCS, the principal plane (or the Galactic equator) coincides
with the plane of the Galaxy and the Galactic longitude of $0^\circ$ is
defined as the direction pointing to the Galactic center (GC).
According to the above criteria, the IAU Sub-Commission 33b summarized
previous observations and defined the GalCS in the framework of the
B1950.0 FK4-based reference system in 1958 (Blaauw et al.
\cite{blaauw60}). The standard GalCS subsequently adopted was that 
for which the pole was
based primarily on the distribution of neutral hydrogen in the inner
parts of the Galactic disk. Murray (\cite{murray89})
discussed the transformation of the GalCS from the FK4 to the FK5
system and derived the orientation of the GalCS at the epoch of
J2000.0, and his method was adopted by the Hipparcos team as a
standard transformation.

Liu et al. (\cite{liu11}) discussed the history of
the GalCS, found that a J2000.0 GalCS which was transformed
from the original IAU1958 definition using unrigorous methods does not match
the physical features of the GalCS, and pointed out that there may then be
some confusions, when applying the Galactic coordinates. Here we
list the transformation matrix and the parameters for the sake of
comparison. The transformation matrix ($3 \times 3$) from the
equatorial to the Galactic coordinate system can be written as
\begin{equation}
\mathcal {N} = \mathcal {R}_3 \left( 90^\circ - \theta \right) \cdot
\mathcal {R}_1 \left( 90^\circ - \delta^{\rm p}\right) \cdot
\mathcal{R}_3 \left( 90^\circ + \alpha^{\rm p} \right),
\end{equation}
where $(\alpha^{\rm p},\delta^{\rm p})$ is the equatorial
coordinates of the north Galactic pole (NGP) and $\theta$ is the
position angle of the Galactic center (GC) at the NGP with respect
to the equatorial pole. These three parameters were defined at the
epoch of B1950.0 by the IAU in 1958 (Blaauw et al. \cite{blaauw60}).
Their numerical values, referred to the FK5-based J2000.0 reference
system, were derived as
\begin{eqnarray}
\alpha_{\rm{J2000.0}}^{\rm{p}}  = 12^{\rm{h}}51^{\rm{m}}26^{\rm{s}}.2755, \nonumber  \\
\delta_{\rm{J2000.0}}^{\rm{p}}  =  +27^\circ07'41''.704,  \nonumber  \\
\theta_{\rm{J2000.0}}  = 122^\circ.93191857, \nonumber \\
i_{\rm{N}} = 90^\circ - \delta_{\rm{J2000.0}}^{\rm{p}} = 62^\circ.8717,
\end{eqnarray}
where $i_{\rm N}$ is the inclination of the Galactic plane.

Starting from the J2000.0 system, Liu et al. (\cite{liu11}) derived the
GalCS in the ICRS by using the bias matrix $\mathcal B$, and the
authors also tentatively proposed a new definition of the GalCS
using the position of the compact radio source Sagittarius A* (Sgr
A*). They only used a single observation of the GC (ICRS position of
Sgr A*) to construct a new coordinate system, failing to using 
other data to ensure that this revised definition
is the optimal choice. Large astrometric
and photometric catalogs such as the Two-Micron All Sky Survey
(2MASS) and other catalogs at radio wavelength such as SPECFIND v2.0
have provided us with deeper and more homogeneous observational data
to estimate more reliably the position of the Galactic plane. 
To construct an optimal GalCS, they also ensured that there was an accurate match with the real Galactic plane on the celestial sphere, namely, that the principal plane
($x-y$ plane) corresponds to the median plane of the Milky Way and
its $x$-axis pointing to the GC.

It is known that the Milky Way is a disk-like Galaxy in which our
Sun is located. The scale height of the thick disk is about 1350 pc
(Gilmore \& Reid \cite{gilmore83}) or 1048 pc (Veltz et al.
\cite{veltz08}). Since the vertical displacement of the Sun above the
Galactic disk ($\sim$15 pc) (Zhu \cite{zhu09}) is much smaller
than the Galactocentric distance of the Sun ($\sim$8.0 kpc)
(Reid \cite{reid93}) and the scale height of the Galaxy, we can
assume that the Sun is in the middle of the Galactic disk so that
the Milky Way is an ``edge-on" galaxy seen from the Sun, which
leads to equal numbers of member stars above and below the mean
Galactic plane. A deep and homogeneous all-sky survey should reveal
the edge-on structure of the Milky Way and the stars should be evenly
distributed above and below a median great circle on the celestial
sphere, which is the optimal Galactic plane we intend to find. In
this paper, we select and analyze data from the latest near-infrared 
and radio catalogs to calculate the orientation of the
optimal GalCS in the ICRS.

In Sect. 2, we describe the properties of the data that we use to
fit the mean position of Galactic plane. Sect. 3 presents the
results for the Galactic plane and the GC positions, and we discuss our revised
definitions of the GalCS in Sect.4. Finally, in Sect.
5 we discuss and summarize our main results and draw some conclusions.

\section{Data}
We attempt to estimate the position of the Galactic plane using data describing
the all-sky distribution of stars. Our goal is to calculate three
parameters $(\alpha^{\rm p},\delta^{\rm p},\theta)$ that describe
the orientation of the GalCS in the ICRS. A deep all-sky catalog
with stars complete to a certain magnitude would reproduce a smooth
Galactic belt on the celestial sphere and a highest star density in
the direction of the GC if we did not consider interstellar
extinction and the vertical distance of the Sun above the Galactic
disk.

A catalog has to meet several criteria if it can be used to
determine reliably the position of the Galactic plane. First, it
must cover the area of the whole disk at low Galactic latitudes.
Secondly, it must be affected by little or no interstellar extinction so that
the two dimensional (on the celestial sphere) distribution of the
stars is uniform. Infrared or radio catalogs usually meet this
criteria but optical catalogs are strongly affected especially
at low Galactic latitudes. Thirdly, it should contain a sufficient
number of stars in the framework of the ICRS. The positional
precision of the individual stars are not crucial because we rely on a
statistical least squares fit method (see the next section) to obtain
the most reliable estimate of the mean position of the Galactic plane.
Proper motions of stars are not needed either, because the newly
defined GalCS is designed to be epoch independent and fixed with
respect to the ICRS.

We choose the 2MASS point source catalog (Cutri et al.
\cite{cutri03}, Skrutskie et al. \cite{skrutskie06}) at near
infrared wavelengths and the SPECFIND v2.0 (Vollmer et al.
\cite{vollmer10}) catalog in the radio band which satisfy all of the above
criteria, in our analysis here.

\subsection{2MASS}
The 2MASS catalog was constructed from uniform observations of the
entire sky (covering 99.998\% of the celestial sphere) in three
near-infrared bands: $J$ (1.25\,$\rm{\mu m}$), $H$ (1.65\,$\rm{\mu
m}$), and $K_s$ (2.16\,$\rm{\mu m}$). The limiting flux in each band
is about 1 mJy. The 2MASS contains positions and photometric
parameters of 470 992 970 sources, but no proper motions. The
positions of stars are calibrated to the Tycho-2 catalog and the
astrometric accuracy is 70-80 mas in the magnitude range of
$9<K_s<14$ mag. Figure 13 of Skrutskie et al. (\cite{skrutskie06})
shows a beautiful pictures of our Galaxy across the Sky. The 2MASS
scans have a typical signal-to-noise ratio of ten and is $>99\%$ complete above 15.8,
15.1, and 14.3 magnitude for $J$, $H$, and $K_s$, respectively. We
choose a complete set of data to fit the Galactic equator by limiting 
the data we consider to those for which $11<J<15.8$.

\subsection{SPECFIND v2.0}
The new release of the SPECFIND radio cross-identification catalog
(Vollmer et al. \cite{vollmer10}) contains 3.76 million
cross-identified radio objects, which has about 60\% more sources
than its first release in 2005. The positions of the radio
objects in the SPECFIND v2.0 catalog were derived from 97
catalogs. Figure 1 depicts the projected distribution of objects
in the J2000.0 Galactic coordinates system from which we can see
clearly the Galactic plane, even though the whole catalog is not very
homogeneous. In this sense, only objects at low Galactic latitudes
are selected.

\begin{figure}
\centering
\includegraphics[scale=0.45]{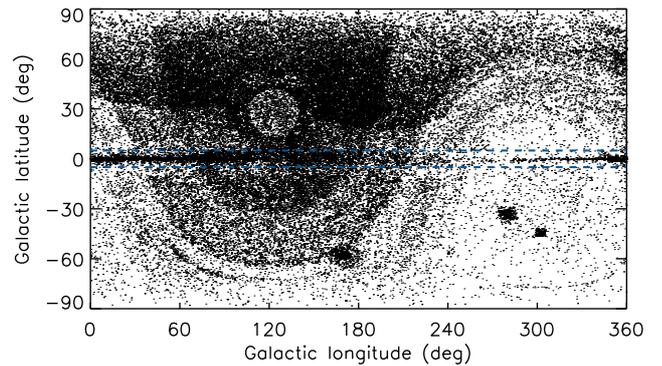}
\caption{The sky distribution of the SPECFIND v2.0 catalog in the
J2000.0 GalCS. The dashed lines are horizontal at $b = -5^\circ$ and
$b = +5^\circ$}
\end{figure}

%

\section{The position of the Galactic plane in the ICRS}
\subsection{Fitting the position of the Galactic plane on the celestial sphere}
The orientation of the Galactic equator can be defined by the
equatorial coordinates of the NGP $(\alpha^{\rm p},\delta^{\rm p})$
or equivalently $(\alpha_{\rm N},i_{\rm N})$, where $\alpha_{\rm N}$
is the right ascension of the ascending node of the Galactic equator
on the celestial equator and $i_{\rm N}$ is the
inclination of the Galactic plane. They are related by
\begin{displaymath}
\alpha_{\rm N} - \alpha^{\rm p} = 90^\circ, \; i_{\rm N} + \delta^{\rm p} = 90^\circ.
\end{displaymath}
Figure 2 shows the relationship between the Galactic plane and the
equator. The equatorial coordinates $(\alpha, \delta)$ of a star
$\sigma$ on the Galactic plane satisfies the equation
\begin{equation}
\tan \delta = \sin \left(\alpha - \alpha_{\rm N}\right)\cdot \tan i_{\rm N},
\end{equation}
in which $\alpha_{\rm N}$ and $i_{\rm N}$ are two parameters to be
fitted from the catalog data.

In this paper, we ignore the weak interstellar extinction at the infrared and
radio bands (which means that the survey observation is homogeneous)
so that the Galaxy has a symmetrical structure as seen from the Sun.
The optimal Galactic equator should be a great circle on the
celestial sphere that is the closest fit to the density distribution
pattern of the above catalogs: the position of the Galactic plane is
where the highest star density appears in the direction of the Galactic
longitudes, because stars should be concentrated on the real Galactic
plane due to the gravitational attraction and rotation of the Milky Way. In other
words, the great circle chosen should have an equal number of stars on both sides.

\begin{figure}
\centering
\includegraphics[scale=0.4]{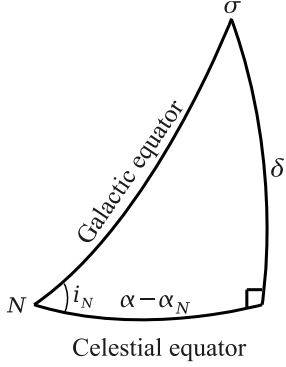}
\caption{The pattern of the Galactic plane and the celestial
equator: $\sigma(\alpha,\delta)$ is a star on the Galactic plane.
$N$ is the ascending node of the Galactic plane with respect to the
celestial equator.}
\end{figure}

For the 2MASS point source catalog, we select stars within the
magnitude range $11<J<15.8$. The upper limit to the magnitude range
is the 99\% completeness magnitude in the $J$ band as mentioned in Sect.
2.1, and the lower limit rejects bright stars close to the
Sun. Owing to the local structure of the of the Galaxy, such as the Gould belt (Westin
\cite{westin85}) and the Galactic warp (Miyamoto et al.
\cite{miyamoto88}), which may affect our results, we exclude nearby stars
by applying a bright upper limit to our selected sample. The complete
sample comprises about 56.7 percentage of stars in the whole 2MASS
catalog. The density map (360 bins in $\ell$ $\times$ 180 bins in
$b$ ) of the selected stars in the J2000.0 Galactic coordinate system is
shown in Fig. 3, which depicts a beautiful edge-on Galactic plane on
the sky. The star density at high Galactic latitudes is much lower than
closer to the Galactic plane and the influence of halo stars is not high.

\begin{figure}
\centering
\includegraphics[scale=0.5]{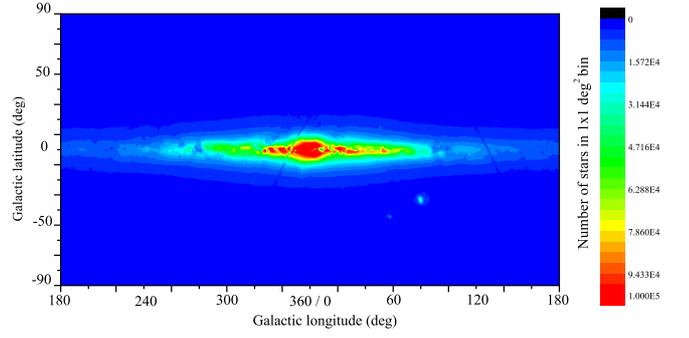}
\caption{The star density distribution of the 2MASS point source
catalog in the magnitude range $11<J<15.8$. The direction of GC
$(\ell=0,b=0)$ is plotted at the center.}
\end{figure}

Instead of fitting Eq. (3) to individual star coordinates
$(\alpha,\delta)$, we choose to use the equatorial positions of the
geometrical centers of each $1^\circ$-Galactic-longitude bin (360
in total), in order to derive a more accurate estimate of $\alpha_{\rm N}$ and
$i_{\rm N}$. We analyze the 2MASS data in this way for the
following reasons. The stellar density close to the GC is extremely high
(which means that the weight of the data in the GC region is much
larger than that in any other area), such that the small-scale
structures close to the GC will strongly affect the resulting position
of the Galactic plane. Our processing method assigns to data at all
Galactic longitudes the same weight to avoid any biasing.
The densities within the LMC and SMC areas are excessively high in the
current study and are replaced with surrounding densities in subsequent calculations.

We first calculate the central Galactic coordinates of the 360
columns of the 360$\times$180 density-distribution matrix and then
transform $(\bar \ell_k,\bar b_k)$, $k=1\cdots360$ into an equatorial
coordinate system to derive the equatorial coordinates $(\bar
\alpha_k,\bar \delta_k)$, $k=1\cdots360$, by applying the previous J2000.0
transformation matrix in Eq. (1). In the next step, these equatorial
coordinates are fitted with Eq. (3) to obtain $\alpha_{\rm N}$
and $i_{\rm N}$ for the first time. With the additional position
angle $\theta_{\rm {J2000.0}}$ in Eq. (2), we can write the new
transformation matrix $\mathcal N$, according to which the updated
Galactic coordinates of the stars and the updated density matrix of
our samples can be obtained. Repeating the above processes, we
derive converged parameters $\alpha_{\rm N}$ and $i_{\rm N}$, which
have the numerical values
\begin{displaymath}
\alpha_{\rm{N}} = 282^\circ.987 \pm 0.032,\; i_{\rm{N}} = 63^\circ.280 \pm 0.019.
\end{displaymath}
These values of $(\alpha_{\rm N}, i_{\rm N})$ are equivalent to the IAU definition
parameters $(\alpha^{\rm p},\delta^{\rm p})$ and we can quote them
to infinite accuracy (here and hereafter to 1 milli-arcsecond), if
they are adopted as a definition
\begin{eqnarray}
\alpha^{\rm{p}}_{\rm{2MASS}}  = 12^{\rm{h}}51^{\rm{m}}56^{\rm{s}}.7726, \nonumber  \\
\delta^{\rm{p}}_{\rm{2MASS}}  =  +26^\circ43'11''.096 .
\end{eqnarray}
It should be understood that this is a numerical convention and does
not necessarily imply that any given parameters and elements of the
matrices (in the following equations) are known to the quoted accuracy.
The standard $1-\sigma$ errors in above results reflect only the uncertainty 
in the data fitting and have no deep physical meanings.

Figure 4 illustrates the positions of the 360 bin centers, in which
the gray curve shows the original value, whereas the black one is plotted
after $\alpha_{\rm N}$ and $i_{\rm N}$ converge. The raw and reduced
curve are almost evenly situated above and below $b=0$. They have
similar shapes along the great circle $b =0$ and correspond to the
old and new transformation matrix $\mathcal N$. The difference between
the two curves reflects the small-angle orientation bias between the
original and the final GalCSs.

\begin{figure}
\centering
\includegraphics[scale = 0.5]{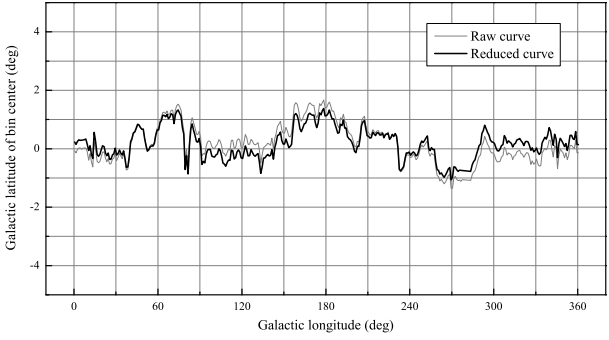}
\caption{The curve of the center in $1^\circ$ longitude bins. The
gray curve is the original curve transformed by J2000.0 rotation
matrix and the black one the reduced curve transformed by new matrix
that is derived from 2MASS fitting.}
\end{figure}

In the SPECFIND v2.0 catalog (see Fig. 1), objects are not
evenly distributed on the sky: the southern hemisphere was observed
less than the northern. The global structure of this distribution 
has a strong effect on our results, even though the Galactic structure can be
clearly seen in Fig. 1. We only retain the data at low
Galactic latitudes $|b|<5^\circ$, which avoids possible biasing. 
We divide the belt $(0^\circ\le l < 360^
\circ,-5^\circ < b < 5^\circ)$ into 360 bins, each one with
$1^\circ(\rm{in}\;\ell) \times 10^\circ(\rm{in}\; b)$ coverage. The
mean positions of objects in each bin $(\bar \ell_{\rm{k}}, \bar
b_{\rm{k}})$, $k=1,\cdots,360$ are then calculated and transformed into
the J2000.0 equatorial coordinates $(\bar \alpha_{\rm{k}}, \bar
\delta_{\rm{k}})$, $k=1,\cdots,360$. Similarly as we process the
2MASS catalog, these 360 equatorial coordinates are used to to fit
$\alpha_{\rm N}$ and $i_{\rm N}$ in Eq. (3) using a least squares method.
After we obtain the resulting $\alpha_{\rm N}$ and $i_{\rm N}$, the
equatorial to Galactic transformation matrix $\mathcal N$ can be
derived by applying $\theta_{\rm{J2000.0}}$. The equatorial
coordinates of the raw data can then be transformed into the Galactic
coordinates using the new matrix. In the next step, we recalculate the
updated coordinates of the mean positions $(\bar \ell_{\rm{k}}, \bar
b_{\rm{k}})$ and fit Eq. (3) to get another pair of $\alpha_{\rm N}$
and $i_{\rm N}$ values. These procedures are repeated until the two
objective parameters converge to the final values. The resulting
parameters are

\begin{displaymath}
\alpha_{\rm{N}} = 282^\circ.699 \pm 0.031,\;\; i_{\rm{N}} = 63^\circ.513 \pm 0.020 ,
\end{displaymath}
or the coordinates of the NGP are
\begin{eqnarray}
\alpha^{\rm{p}}_{\rm{SPECFIND}}  = 12^{\rm{h}}50^{\rm{m}}47^{\rm{s}}.6503  \nonumber  \\
\delta^{\rm{p}}_{\rm{SPECFIND}}  =  +26^\circ29'12''.412 .
\end{eqnarray}

We note that the position angle $\theta_{\rm {J2000.0}}$
(which indicates the position of the GC on the Galactic plane) is
unaltered in iterative steps because we are only concerned with the position
of the Galactic plane in both the 2MASS and SPECFIND v2.0 data sets.
From Eqs. (4) and (5), we find that the fitted values of
inclination $i_{\rm N}$ from the 2MASS and the SPECFIND v2.0 are
about $0.41^\circ$ and $0.64^\circ$ respectively, larger than the
J2000.0 reference values in Eq. (2). The ascending node fitted using
the 2MASS data is about 7 arcseconds eastward of the J2000.0 node,
but the corresponding result using the SPECFIND v2.0 data is about 10 arcseconds 
westward of the J2000.0 node. We emphasize here that our results depend on the choice
of data (range of magnitude, type of stars, bandwidths, etc.),
although both 2MASS and SPECFIND v2.0 are appropriate candidate catalogs
to fit $\alpha_{\rm N}$ and $i_{\rm N}$ because of their advantages
described above. The generated results in Eqs. (4) and (5) present, to a
certain extent, the position of the Galactic plane in the near-infrared
and radio continuum data sets based on modern observations. We
confirm that complete and deeper long-wavelength surveys will
provide more reliable estimates. Although there are many extremely large optical
catalogs, such as UCAC3 (Zacharias et al. \cite{zacharias10}),
USNO B1.0 (Monet et al. \cite{monet03}), and the latest PPMXL catalog
(Roeser et al. \cite{roser10}) (of nearly one billion stars), they cannot 
be used in this study, mainly because of the non-isotropic
interstellar extinction along different lines of sight, which destroys the
homogeneity of the density distribution of stars in these optical
catalogs.

\subsection{The position of the Galactic center in the ICRS}

The IAU working group reviewed the precise position of the compact radio
source Sagittarius A*, in the B1950.0 reference system (Gum \& Pawsey
\cite{gum60}) to be
\begin{displaymath}
\alpha^0_{\rm {B1950.0}} = 17^{\rm h}42^{\rm m}37^{\rm s}, \;
\delta^0_{\rm {B1950.0}} = -28^\circ 57',
\end{displaymath}
which was afterward used to define the direction of the $x$-axis (GC)
of the IAU1958 GalCS. According to the rigorous FK4-FK5
transformation proposed by Standish (\cite{standish82}) and Aoki et
al. (\cite{aoki83}), the position of the Sgr A* at J2000.0 is
\begin{equation}
\alpha^{\rm {Sgr A*}}_{\rm {J2000.0}} = 17^{\rm h}45^{\rm m}47^{\rm s}.7, \;
\delta^{\rm {Sgr A*}}_{\rm {J2000.0}} = -28^\circ.58'09'',
\end{equation}
while the $x$ axis direction of the J2000.0 GalCS recommended by
Hipparcos team (ESA \cite{esa97}) is
\begin{equation}
\alpha^0_{\rm {J2000.0}} =17^{\rm h}45^{\rm m}37^{\rm s}.1991, \;
\delta^0_{\rm {J2000.0}} =-28^\circ56'10''.221.
\end{equation}
Using Very Long Baseline Array (VLBA) measurements, Reid \&
Brunthaler (\cite{reid04}) derived the absolute position (referred
to the ICRS) of the Sgr A* at the dynamical center of the Galaxy
\begin{equation}
\alpha^{\rm {Sgr A*}}_{\rm {Reid}} =17^{\rm h}45^{\rm m}40^{\rm s}.0400, \;
\delta^{\rm {Sgr A*}}_{\rm {Reid}} =-29^\circ00'28''.138.
\end{equation}
Since the Milky Way is a gravitational system centered at the GC,
the stellar density should be highest in that region. We find from
the 2MASS data set that the position of the peak density is
\begin{equation}
\alpha^0_{\rm {2MASS}}= 17^{\rm h}46^{\rm m}6^{\rm s}.5328, \;
\delta^0_{\rm {2MASS}}= -28^\circ54'16''.164.
\end{equation}
In the following section, we adopt the Sgr A* position of Reid \&
Brunthaler (\cite{reid04}) derived using direct radio observations as the
direction of the GC because it is independent of complicated
transformations.
%

\section{Constructing the Galactic coordinate system}
We denote the orthogonal GalCS using three principal axes $[x,y,z]$.
The orientation of the IAU1958 and Hipparcos (J2000.0) GalCSs are
characterized by the coordinates of the NGP ($z$-axis) and the
position angle of the GC ($x$-axis). After this set of parameters has been
determined, we can construct the GalCS in the corresponding equatorial
reference system using the $3 \times 3$ matrix $\mathcal N$. In Sect. 3,
we have derived possible the positions of the NGP (Galactic
plane) $(\alpha^{\rm p},\delta^{\rm p})$ and the GC
$(\alpha^0,\delta^0)$. However, these coordinates or angles cannot
be used to construct the GalCS directly, because their directions in
$(\alpha^{\rm p},\delta^{\rm p})$ and $(\alpha^0,\delta^0)$ were
determined independently, either by catalog fitting or by observations, and they are not
perpendicular to each other. In other words, a given GC may not lie
on a given Galactic plane. We construct the GalCS in two different
ways: (1) we assume that the $z$-axis coincides with the given direction
of the NGP $(\alpha^{\rm p},\delta^{\rm p})$ then try to find the
$x$ direction using the given GC coordinates $(\alpha^0,\delta^0)$;
(2) we assume that the $x$-axis coincides with the given direction
$(\alpha^0,\delta^0)$ of the GC then try to infer the
orientation of the Galactic plane from catalog fitting.

\subsection{Method 1: $z$-axis fixed condition}
We first assume that the $z$-axis of the GalCS points in the
direction of the NGP ($\alpha^z=\alpha^{\rm
p},\;\delta^z=\delta^{\rm p}$) that is fitted from either the 2MASS or the
SPECFIND v2.0 catalog. Since the given GC does not lie on the
Galactic plane ($x$-$y$ plane), we try to find the nearest point
$(\alpha^x, \delta^x)$ from the given GC $(\alpha^0,\delta^0)$ on
the fixed Galactic plane $(\alpha^{\rm p},\delta^{\rm p})$.
Evidently, $(\alpha^x, \delta^x)$ and $(\alpha^0,\delta^0)$ share a
common position angle $\theta$ with respect to the north celestial
pole as seen from the NGP, given by
\begin{equation}
\tan \theta=\frac
{\sin \left( \alpha^0 - \alpha^{\rm p} \right)}
{\cos \delta^{\rm p} \cdot \tan \delta^0
- \sin \delta^{\rm p} \cdot \cos \left( \alpha^0 - \alpha^{\rm p} \right)}.
\end{equation}
Substituting in both Eqs. (4) and (8), we find that
\begin{equation}
\theta_{\rm {2MASS}} = 123^\circ.044663150.
\end{equation}
On the basis of these values of $\alpha^{\rm p}_{\rm {2MASS}}$,
$\delta^{\rm p}_{\rm {2MASS}}$, and $\theta_{\rm {2MASS}}$, we derive
our new equatorial-Galactic transformation matrix based on the 2MASS data set
and the Sgr A* position (printed to 10 decimals, 1 milli-arcsecond accuracy)
\begin{eqnarray}
&&\mathcal N_{\rm{2MASS}}= \nonumber \\
&&\left(
\begin{array}{rrr}
-0.0505347007 &    -0.8719028362 &    -0.4870643574  \\
+0.4897972973 &    -0.4466482209 &    +0.7487349160  \\
-0.8703705255 &    -0.2007257110 &    +0.4496268868
\end{array}
\right).
\end{eqnarray}

If we adopt Eqs. (5) and (8) to compile the GalCS, the position angle of
$x-$axis and the rotation matrix are
\begin{equation}
\theta_{\rm {SPECFIND}} = 122^\circ.914360525,
\end{equation}
and
\begin{eqnarray}
&&\mathcal N_{\rm{SPECFIND}}= \nonumber \\
&&\left(
\begin{array}{rrr}
-0.0518807421 &    -0.8722226427 &    -0.4863497200  \\
+0.4846922369 &    -0.4477920852 &    +0.7513692061  \\
-0.8731447899 &    -0.1967483417 &    +0.4459913295
\end{array}
\right).
\end{eqnarray}

\subsection{$x$-axis fixed condition}
In the above calculations, the $z$-axis of a constructed GalCS coincides
perfectly with the fitted NGP, but the $x-$axis direction does not
match the given GC (Sgr A*), but instead offset by several arcminutes. 
We now assume that the $x$-axis coincides with the given GC
$(\alpha^x=\alpha_0,\;\delta^x=\delta_0)$, and
determine the coordinates of the NGP ($z$ direction) by fitting either the
2MASS or SPECFIND v2.0 catalogs with the same strategy as in Sect. 3.
Since we have already determined two of the three parameters that define the
GalCS, there remains only one parameter to be determined. We choose
the position angle $\eta$ of the NGP (seen from the given GC) as the
last unknown parameter. We refer to Fig. 5 for the definition of $\eta$. The
J2000.0 reference value of $\eta$ is
\begin{displaymath}
\eta_{\rm {J2000.0}} = 58^\circ.599,
\end{displaymath}
and the equation for fitting is then
\begin{equation}
\cos \delta^0 \cdot \tan \delta = \sin \delta^0 \cdot \cos \left(
\alpha - \alpha^0 \right) + \sin \left( \alpha - \alpha^0 \right)
\cdot \tan \eta.
\end{equation}

\begin{figure}
\centering
\includegraphics[scale=0.4]{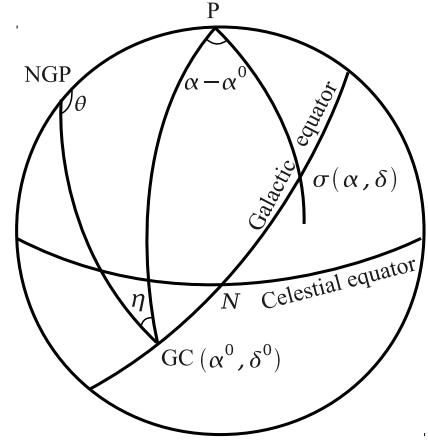}
\caption{The position angle $\eta$ of the NGP at the given GC
$(\alpha^0,\delta^0)$, where $\theta$ is the position angle of the GC,
$\sigma(\alpha, \delta)$ is a star on the Galactic equator, and $P$
is the celestial pole.}
\end{figure}

The equatorial coordinate system to GalCS transformation matrix
using $(\alpha^0, \delta^0, \eta)$ has the form
\begin{equation}
\mathcal {N} = \mathcal {R}_1 \left(\eta \right) \cdot \mathcal
{R}_2 \left( -\delta^0 \right) \cdot \mathcal{R}_3 \left( \alpha^0
\right).
\end{equation}
With the matrix $\mathcal N$, three standard parameters $\alpha^{\rm
p}$, $\delta^{\rm p}$, and $\theta$ of the IAU scheme can be
determined. In particular, we specify the coordinates in Eq. (8) to
be the $x-$axis and fit the 2MASS stellar density data iteratively
with Eq. (15) to determine $\eta$ and $\mathcal N$. The final
results are
\begin{equation}
\eta_{\rm{2MASS}} = 58^\circ.947 \pm 0.025,
\end{equation}
\begin{eqnarray}
&&\mathcal N_{\rm{2MASS}}  = \nonumber \\
&&\left(
\begin{array}{rrr}
-0.0546572359 &    -0.8728439269 &    -0.4849289286  \\
+0.4888603641 &    -0.4468595864 &    +0.7492209651  \\
-0.8706481098 &    -0.1961121855 &    +0.4511229097
\end{array}
\right),
\end{eqnarray}
\begin{eqnarray}
&&\alpha^{\rm{p}}_{\rm{2MASS}}  = 12^{\rm{h}}50^{\rm{m}}46^{\rm{s}}.5444, \nonumber  \\
&&\delta^{\rm{p}}_{\rm{2MASS}}  =  +26^\circ48'56''.706, \nonumber \\
&&\theta_{\rm{2MASS}} = 122^\circ.912729244.
\end{eqnarray}

Similarly, we obtain the orientation of the GalCS based on Eq. (8)
and the SPECFIND v2.0 catalog
\begin{equation}
\eta_{\rm {SPECFIND}} = 59^\circ.275 \pm 0.024,
\end{equation}
\begin{eqnarray}
&&\mathcal N_{\rm{SPECFIND}}  = \nonumber \\
&&\left(
\begin{array}{rrr}
-0.0546572359 &    -0.8728439269 &    -0.4849289286  \\
+0.4838685275 &    -0.4479748647 &    +0.7517910405  \\
-0.8734322153 &    -0.1935510264 &    +0.4468267735
\end{array}
\right),
\end{eqnarray}
\begin{eqnarray}
&&\alpha^{\rm{p}}_{\rm{SPECFIND}}  = 12^{\rm{h}}49^{\rm{m}}58^{\rm{s}}.7360 \nonumber  \\
&&\delta^{\rm{p}}_{\rm{SPECFIND}}  =  +26^\circ32'24''.989 \nonumber \\
&&\theta_{\rm{SPECFIND}} = 122^\circ.823292026.
\end{eqnarray}
The resulting values of $\eta$ from the both 2MASS and the SPECFIND v2.0 differs by a few
tenths of a degree when referred to $\eta_{\rm {J2000.0}}$.

The GalCSs derived by applying the two methods with the same observational
data (2MASS or SPECFIND) differ in terms of the small-angle orientation by
several arcminutes, while the bias in the GalCSs caused by the use of different
observational data is about a few tenths of an arcsecond. In Sect. 4 of Liu
et al. (\cite{liu11}), the authors proposed a possible definition of
the GalCS based on the ICRS position of Sgr A* and the transformed
GalCS, while the present proposals depend only on the observational data but
are unrelated to the old GalCS and its transformations between various
fundamental reference systems, which should be avoided when defining a
new coordinate system. In the context of our aim to
establish a revised GalCS, we have succeeded in defining an optimal GalCS 
in which the $z$-axis coincides with the NGP
and the $x$-axis with the position of Sgr A*, which is regarded as the GC. All data
have been used were measured in the framework of the ICRS. The GalCS should be independent
of epoch, thus the transformation matrix should be applied at
the epoch of J2000.0. The results presented in Eqs. (8), (11), (13), (19), 
and (22) provide possible references for future revisions of the definition of the GalCS.

%
\section{Discussion and conclusion}
On the basis of near-infrared and radio observational data, we have described
several approaches for calculating the positions of the Galactic
plane and the GC. From our fitted results, we have found that the inclination
of the Galactic plane with respect to the ICRS equator is either
$0.4^\circ$ (2MASS) or $0.6^\circ$ (SPECFIND v2.0) larger than
commonly used values recommended by the Hipparcos. This implies
that there remains room for improvement in the J2000.0 GalCS.

As mentioned at the beginning of the paper, the main purpose of establishing a robust
GalCS is to ensure that in studies of Galactic kinematics, dynamics, and
structure, the Galactic coordinates $(\ell,b)$ of an object are
always reliably and universally transformed from its equatorial coordinates $(\alpha,
\delta)$. We note that the establishment of the GalCS need not
have the same degree of precision as either the ecliptic or equatorial
reference systems. The FK4, FK5, and Hipparcos or the ICRS are
reference systems based on carefully determined positions of the
equator and equinox, while the Galactic coordinate system is based on
the concentration of tens of millions of stars or continuum radio
objects in a statistical sense. We have defined in the previous sections
a transformation method $(\alpha^{\rm p}, \delta^{\rm p},
\theta)$ that is the most appropriate choice for Galactic studies,
at least at near-infrared and radio wavelengths. This transformation
definition should be specified in certain fundamental reference
systems, the ICRS. Our philosophy is similar to the IAU1958
method and the results are given for two bands, both based on the
latest observations. The parameters presented in Sect. 4 provide some
possible formal definitions of the GalCS that may be applied in the future.

With the development of astrometry to an unprecedented accuracy and the acquisition of a
large amount of data in recent years, we recommend that the Galactic
coordinate system be defined based on observational data
rather than a transformation of the original definition. The optimal definition of
GalCS remains an active field of research that should be considered by relevant IAU
commissions.
%

\begin{acknowledgements}
This work was funded by the National Natural Science Foundation of
China (NSFC) under No. 10973009 and made use of the SIMBAD database
operated at CDS, Strasbourg, France. We are grateful to Dr. Yi Xie
for his careful correction of the language.
\end{acknowledgements}


\end{document}